\documentclass[a4paper,fleqn]{cas-dc}

\usepackage[numbers]{natbib}

\newcommand{\isotope}[2]{\ensuremath{^{#2}\text{#1}}\xspace} 

\newcommand{\isou}{\isotope{U}{238}}
\newcommand{\isoth}{\isotope{Th}{232}}
\newcommand{\isorn}{\isotope{Rn}{220}}
\newcommand{\isorna}{\isotope{Rn}{222}}
\newcommand{\isobi}{\isotope{Bi}{212}}
\newcommand{\isopb}{\isotope{Pb}{212}}
\newcommand{\isopo}{\isotope{Po}{212}}

\usepackage{graphicx} 	
\usepackage{amssymb} 	
\usepackage{mathtools} 	
\usepackage{comment} 	
\usepackage{multirow}	
\usepackage{lineno}

\usepackage[dvipsnames]{xcolor}

\usepackage{xspace} 

\usepackage[shortlabels]{enumitem}

\usepackage[framemethod=TikZ]{mdframed}
\usepackage{xcolor}
\usepackage{empheq}
\usepackage[most]{tcolorbox}

\usepackage{cleveref}

\def\tsc#1{\csdef{#1}{\textsc{\lowercase{#1}}\xspace}}
\tsc{WGM}
\tsc{QE}
\tsc{EP}
\tsc{PMS}
\tsc{BEC}
\tsc{DE}

\begin{document}
\let\WriteBookmarks\relax
\bibliographystyle{unsrt}
\def\floatpagepagefraction{1}
\def\textpagefraction{.001}
\shorttitle{water exrtraction system for JUNO}
\title [mode = title]{Development of water extraction system for liquid scintillator purification of JUNO}                      


\author[1,2]{Jiaxuan Ye}
\cormark[1]
\fnmark[1]
\ead{yejx@ihep.ac.cn}

\author[1,3]{Jian Fang}
\cormark[1]
\ead{fangj@ihep.ac.cn}
\author[1,3]{Tao Hu}
\author[1,3]{Li Zhou}
\author[1,3]{Wei Hu}
\author[1,3]{Wanjin Liu}
\author[1]{Yayun Ding}
\author[1]{Meangchao Liu}
\author[1,3]{Boxiang Yu}
\author[1,3]{Xilei Sun}
\author[1,3]{Lijun Sun}
\author[1,3]{Yuguang Xie}
\author[1,3]{Xiao Cai}

\address[1]{Insitute of High Energy Physics, Chinese Academy of Sciences (CAS), Beijing 100049, China}
\address[2]{University of Chinese Academy of Sciences, Beijing 100049, China}
\address[3]{State Key Laboratory of Particle Detection and Electronics, (Beijing 100049, HeFei 230026), China}

\cortext[cor1]{Corresponding author at: Institute of High Energy Physics, Chinese Academy of Sciences, Beijing 100049, China.}

\begin{abstract}
The Jiangmen Underground Neutrino Observatory (JUNO) uses 20k tons of liquid scintillator (LS) to detect neutrinos. The content of radioactive substances in the liquid scintillator will affect the experimental results. JUNO will use counter current water extraction to reduce the radioactive metal ions inside the LS. In this article, The factors that affect the final water extraction like the partition coefficient and the factors that affect it, the optimal mass transfer droplet size, the flow rate ratio, theoretical stage and different working mode (LS as continuous phase or dispersed phase) have been studied. We built counter current extraction prototype in the laboratory and a pilot plant in Daya Bay. We not only study the factors above, but also get a lot of engineering experience that worth sharing.
\end{abstract}



\maketitle

\section{Introduction}
The Jiangmen Underground Neutrino Observatory (JUNO) is a further exploration of neutrino after the Daya Bay Reactor Neutrino Experiment \cite{juno_cdr}. JUNO is a multi-purpose neutrino observatory. The main goal is to determine the neutrino mass hierarchy as well as to precisely measure neutrino oscillation parameters \cite{juno_white_paper}.

JUNO plans to use 20k tons liquid scintillator(LS) as sensitive medium with an unprecedented target energy resolution of 3\% at 1 MeV. The LS in JUNO is compound organic liquid, containing linear alkyl benzene(LAB)  as primary solvent, 2,5-diphenyloxazole (PPO)  as luminous matter, and tiny amount of 1,4-bis(2-methylstyryl)benzene (bis-MSB) as wavelength shifter \cite{Abusleme:2020bbm}. LAB-based LS has high flash point so is safe to handle. JUNO also has stringent requirements on the radio-purity of the LS in order to reach different physics potentials. For example, in order to study reactor neutrinos, the LS background would be $^{238}$U and $^{232}$Th less than $1\times10^{-15}$ g/g and $^{40}$K less than $1\times10^{-16}$ g/g. Besides, in order to study solar neutrinos, the LS background would be $^{238}$U and $^{232}$Th less than $1\times10^{-17}$ g/g and $^{40}$K less than $1\times10^{-18}$ g/g \cite{juno_white_paper}. Otherwise, these backgrounds will affect the detector’s ability to distinguish neutrino events. Thus the critical point is to reduce the radioactive background in the process of LS production and purification. 

Other large-scale underground neutrino experiments using LS as detection medium, such as Borexino \cite{Alimonti:2000xc} in Italy, SNO+ \cite{Ford:2015zta} in Canada and KamLAND in Japan \cite{Abe:2008aa}, all have similar but different purification methods. Due to the enormous amount of LS, it presents a specific challenge for JUNO to achieve similar level of low background, but with $>$20 times higher LS to purify. JUNO plans to use multiple methods to remove radioactive impurities from LS: distillation, water extraction and gas stripping. Distillation can separate low volatile impurities from LS raw materials \cite{Lombardi:2019epz}. Water extraction is the method that using the solubility difference of metal ions between water and LS, to absorb radioactive metal ions in LS. Gas stripping is to remove radioactive gases in LS, like Ar, Kr, Rn, etc \cite{Lombardi:2019epz}.
As an important purification method, water extraction does not change the LS component, thus can be applied as an online purification method during neutrino experiment running. 


\section{Principle of water extraction}
\label{sec:principle}

Water extraction is a type of liquid-liquid extraction, which is often used to extract solute in one liquid between two immiscible liquids based on the different solubility. In JUNO, Water extraction is used to remove radioactive impurities, mainly $^{238}$U, $^{232}$Th and $^{40}$K, in organic LS. Polar heavy metal ions have large solubility in polar water than non-polar LS \cite{Leung:2006kk}.

Water has many advantages as an extractant. Water is compatible with JUNO LS, besides, it is cheaper and can be used in large quantities since JUNO has 20k tons’ LS. The waste water is easy to handle and has no impact on the environment.

In JUNO, water extraction system will run in counter-current mode. Large industrial system often employs column extraction with multiple theoretical stages rather than single batch extraction to achieve more efficient separation. There are many factors that will affect the extraction efficiency, such as the height of the tower, additional energy (like stirring), the design of the tower internals, the flow rate of water($F_W$) and the flow rate of LS($F_{LS}$), and the temperature during extraction. These factors will affect the theoretical stages of extraction. Theoretical stages is a mathematical description, which means that the extraction reaches equilibrium. 

The partition coefficient $K_D$ is a key parameter to determine the water extraction efficiency of the system. It is defined as the ratio of the solute concentration in water phase ($C_{w}$) divided by the concentration in LS phase ($C_{LS}$) after reaching one theoretical stage. $K_D$ is a function of multiple factors, such as temperature, pressure and ion species. In addition, the choice of continuous phase and dispersed phase may also affect the final efficiency in column extraction. 
\begin{equation}
\label{eq:pc}
    K_{D}=\frac{C_W}{C_{LS}}
\end{equation}

\begin{figure}
    \centering
    \includegraphics[width=1\linewidth]{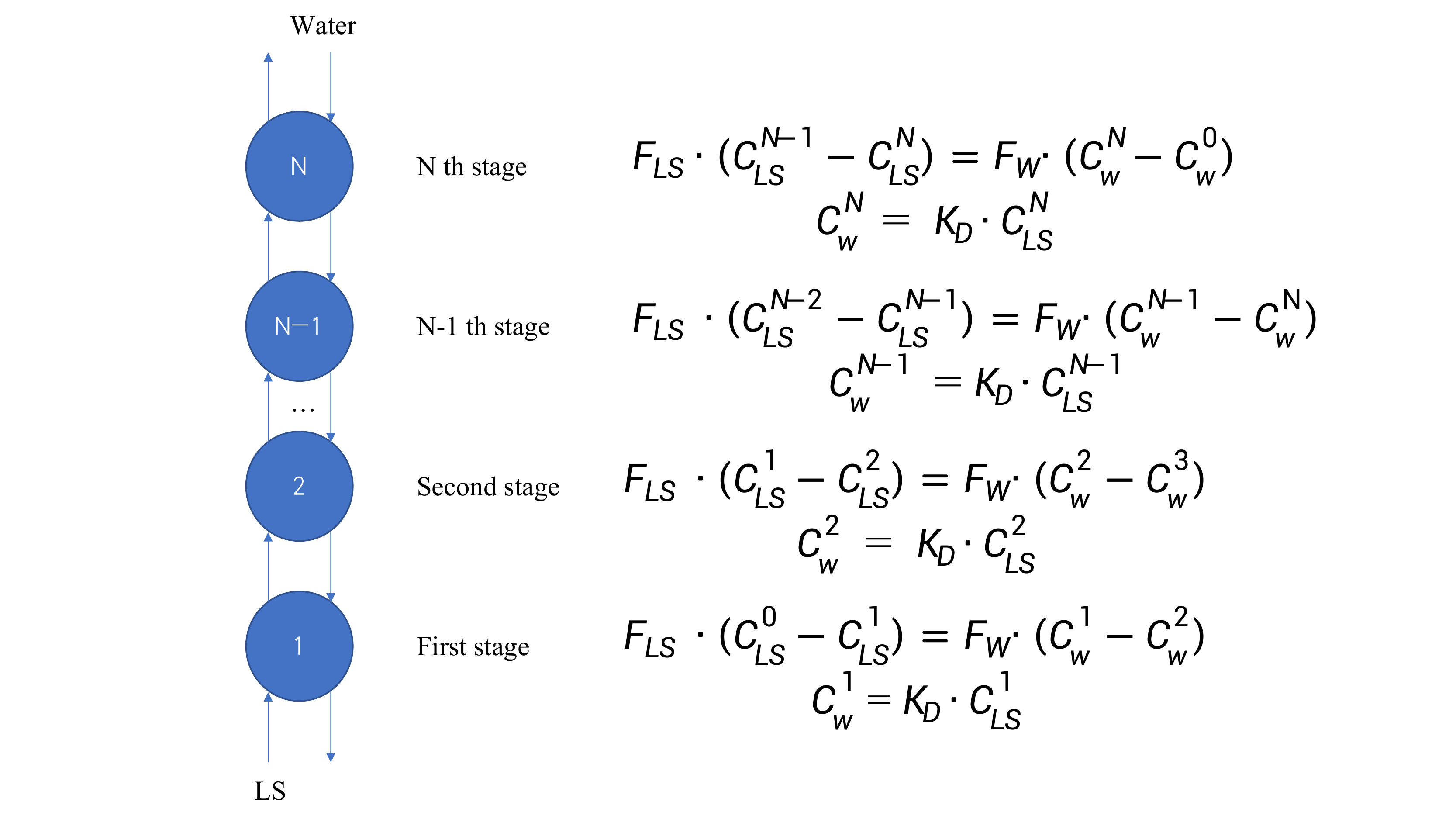}
    \caption{absorption spectrum after four system}
    \label{fig:absorption spectrum}
\end{figure}

Fig. \ref{fig:absorption spectrum} is a schematic diagram of counter-current extraction. ${C_{W}^{0}}$ and ${C_{LS}^{0}}$ represents the concentration of solute in the initial water phase and LS phase. The numbers 1, 2, ..., N in the picture represent the theoretical stages. ${C_{LS}^{N}}$ represents the concentration of solute after Nth stages of extraction. For each theoretical stages, two formulas can be listed according to material balance and definition of partition coefficient. 
Add the formulas of N material balance together to get eq.~\eqref{eq:add}:
\begin{equation}
\label{eq:add}
    C_{LS}^{0}-C_{LS}^{N}=\frac{F_{W} \cdot 
    {K_D}}{F_{LS}}\cdot\left(C_{LS}^{1}-\frac{C_{W}^{0}}{K_{D}}\right)
\end{equation}
After replacing the right side of each material balance formula with the partition coefficient formula, N formulas iterate can be obtained eq.~\eqref{eq:ite}:
\begin{equation}
\label{eq:ite}
    C_{LS}^{0}-C_{LS}^{1}=\left(\frac{F_{W} \cdot K_{D}}{F_{LS}}\right)^{N}\cdot\left(C_{LS}^{N}-\frac{C_{W}^{0}}{K_{D}}\right)
\end{equation}
Using formulas \eqref{eq:add} and \eqref{eq:ite}, the concentration of solute in the LS(${C_{LS}^{N}}$) after N-stage counter-current extraction can be obtained as
\begin{equation}
\label{eq:finaleff}
    C_{LS}^{N}=\frac{\left(1-\frac{F_{LS}}{F_W\cdot K_{D}}\right)\cdot \left(C_{LS}^{0}-\frac{C_{w}^{0}}{K_{D}}\right)}{\left(\frac{F_W \cdot K_{D}}{F_{LS}}\right)^{N}-\frac{F_{LS}}{F_W\cdot K_{D}}}+\frac{C_{W}^{0}}{K_{D}}
\end{equation}

The concentration of solute in the final LS (${C_{LS}^{N}}$) depends on the number of theoretical stages, the partition coefficient, and the ratio of the flow rate of LS to water. In addition, the initial solute concentration of water also have effects.

The goal for this development program is to measure the factors above and gain experience in order to guide the design of full-scale system for JUNO based on the results after experiments.

\section{Laboratory study}
\label{sec:lab_test}

Much efforts have been put to understand the purification efficiency for JUNO LS purification. As the input to water extraction is already distilled, the content of various radioactive metal ions is expected to be at trace level, and is far below the saturated solubility. 

\subsection{General method}
\label{sec:method_description}

The typical \isou/\isoth contamination in JUNO LS are from ${10^{-13}}$ to ${10^{-14}}$ g/g. It is impractical to measure \isou/\isoth content less than $10^{-14}$ g/g at laboratory level \cite{Hu:2016jwc}. Instead of knowing the water extraction efficiency of \isou/\isoth, we adopt the method of using \isopb an proxy to evaluate the water extraction efficiency \cite{Quirk:2008zz}. The method for loading \isopb into LS and setup is descried elsewhere \cite{Hu:2016jwc}. Here we give a brief review.

Our initial goal is to study the extraction efficiency of \isou/\isoth, and then obtain the factors like partition coefficient of \isou/\isoth to guide the final full scale water extraction system design, but a large detector is needed for measurement, so a pilot plan is required. The design of the pilot plant also requires partition coefficient. Since the solubility of uranium and thorium salt compounds in water is higher than that of lead salt compounds\cite{2016CRC}, it is assumed that the partition coefficient of uranium and thorium is not lower than that of lead. Therefore, \isopb extraction efficiency can be used to guide the design of pilot plant. On the other hand, \isopb has a short half life of 10.6 h, allowing it to be measured easier than \isou/\isoth through decay products. We use \isorn to load \isopb into LS. \isopb is the longest lived daughter product through the \isorn decay chain. The number of \isopb decays can be measured via the correlated $\beta-\alpha$ emission from \isobi-\isopo cascade decay. The short characteristic halflife of 300 ns is extremely useful to suppress background. Because of the short half-life, \isopb loading at $10^{-18}$ g/g level could still be measured through their decay products. To set the scale, \isopb loading at $10^{-17}$ g/g in 100 ml of LS gives a decay rate of $\sim$40 Bq.

Continuously fed Rn220 into the LS bubble bottle to 74.2 hours(7 half-life of \isopb), \isopb concentration in LS achieved balance. In the experiments of this paper LS all had loaded \isorna for 74.2 hours in advance. The LS sample is placed in a cylindrical quartz bottle, which is further placed into a dark box. Two 2-inch PMTs (XP2020, HZC) were mounted on each side of quartz bottle. Quartz bottle was 5 cm in diameter and 1 cm in thickness, containing 17.1 g LS. Around the PMTs and LS bottle placed 5cm in thickness low-radioactivity lead bricks to shield part of environment gamma and cosmic ray. 

A CAEN Flash ADC DT5751 (10 bit, 1 GHz sampling rate) was used for data acquisition. A coincidence between two PMTs is required for trigger. Once triggered, event waveform with total length of 2000 ns is recorded. Offline analysis select candidate events with two pulses separated within 5 ns. The total background event rate is 0.26 Hz. 
\subsection{Benchtop test}
\label{sec:benchtoptest}

To measure the partition coefficient of \isopb between water and LS, we use magnetic stirrer, 100 ml Erlenmeyer flask and 3 cm PTFE coated magnet to perform water extraction. \isopb loaded LS is mixed with high-purity water in equal amount (30 ml). The magnetic stirring is to make LS and water to fully mixed and speed up the equalibrium. \isopb rate is measured before ($R_0$) and after ($R_t$) the extraction, with the method described in Sec:~\ref{sec:method_description} . The purification efficiency is 
\begin{eqnarray}\label{eq:distri}
    \epsilon = 1-\frac{R_t}{R_0 e^{- \frac{\Delta t}{\tau}}} = \frac{K_D}{K_D+1}
\end{eqnarray}

\noindent where $\Delta T$ is the time difference between measurements, $\tau$ is the lifetime of \isopb. $K_D$ is the partition coefficient defined in Eq.~\eqref{eq:pc}.

\begin{figure}[h!]
    \centering
    \includegraphics[width=0.95\linewidth]{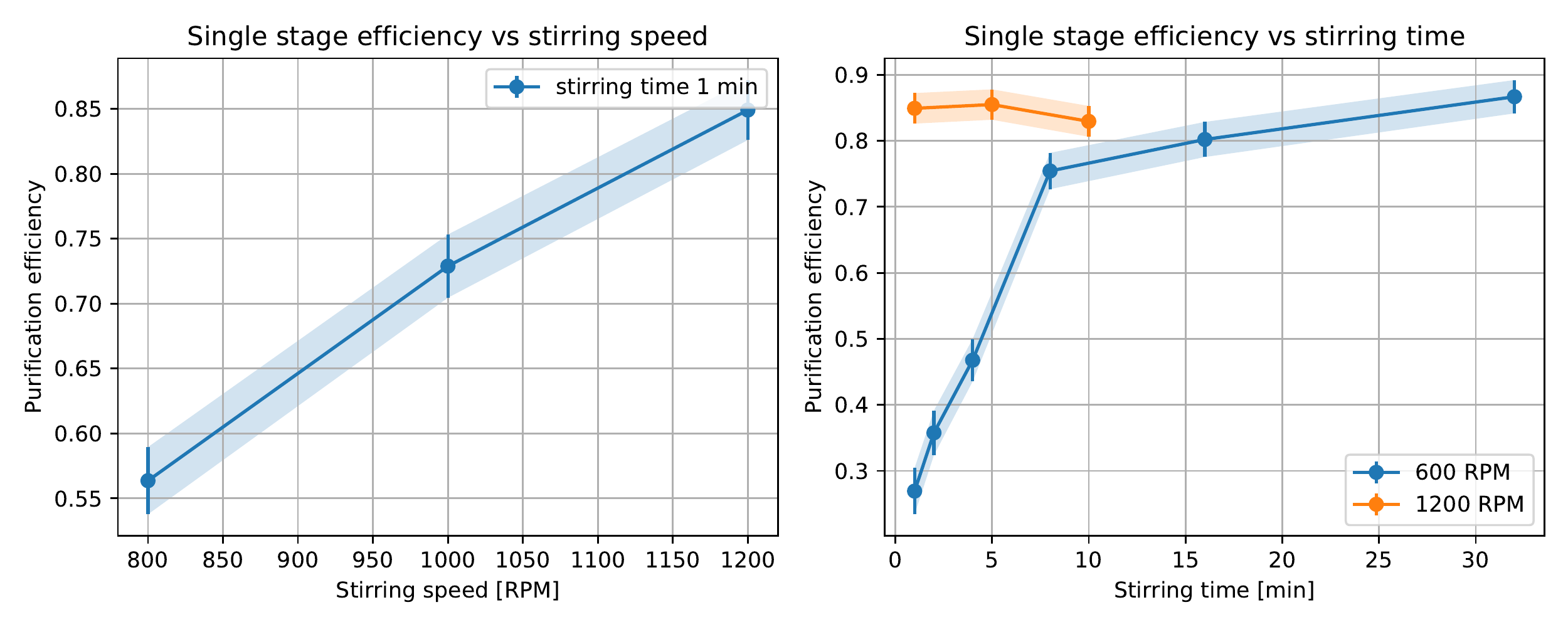}
    \caption{Single stage extraction efficiency with different stirring speed and time. }
    \label{fig:batch_eff_vs_time}
\end{figure}

The measurement of $K_D$ replies heavily on how well the system are in equilibrium. Equilibrium depends on the stirring speed and the total time. With this magnetic stirring, the longer and the faster the stirring, the easier the system reaches equilibrium. Fig.~\ref{fig:batch_eff_vs_time}, the extraction efficiency increases with stirring speed and time before reaching equilibrium. At higher speed, less time are needed to reach equilibrium. It took at least 30 minutes to reach equilibrium at 600 rpm, while only several minutes at 1200 rpm. The single stage extraction efficiency can reach 85$\pm$2\%, with the uncertainty dominated by systematic variations in equilibrium values under different stirring time/speed combinations. 2\% is the systematic error estimated by repeated measurements. This incidates that the partition coefficient of \isopb is from 4.88 to 6.69.

\begin{figure}[h!]
    \centering
    \includegraphics[width=0.8\linewidth]{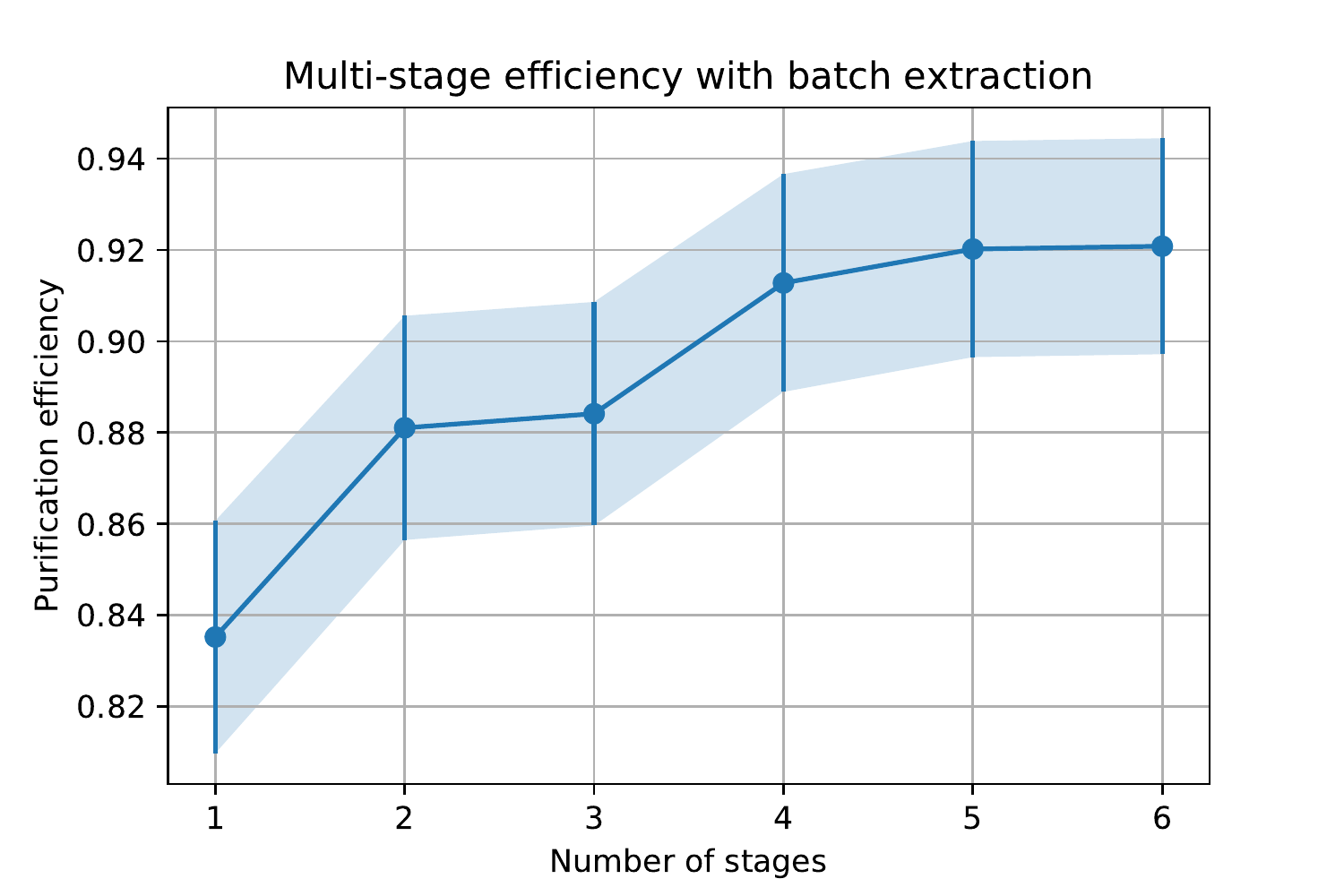}
    \caption{Efficiency and number of extraction stages. Each stage consists of an operation of 3 min stirring at 1200 rpm.}
    \label{fig:eff_vs_stage}
\end{figure}

Multi-stage extraction is also performed. Fig.\ref{fig:eff_vs_stage} shows the relation between extraction efficiency and the number of extraction stages. In this test, each stage consists of 3-min stirring at a speed of 1200 rpm. After 5 stages,  the extraction efficiency became steady, reaching 92\%. This indicates part of the \isopb is in different form that can not be removed by water. 

\begin{figure}[h!]
    \centering
    \includegraphics[width=0.8\linewidth]{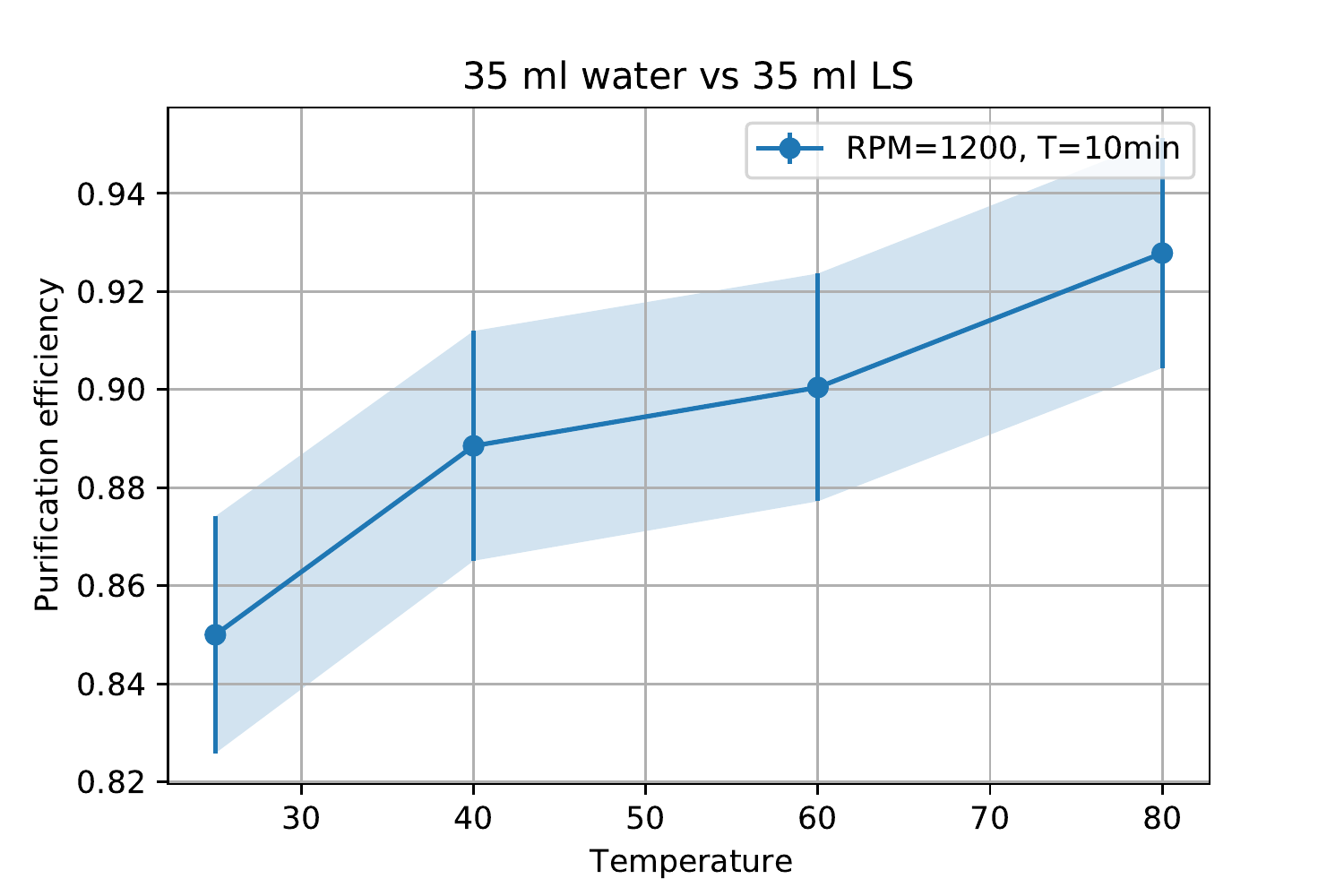}
    \caption{Efficiency vs working temperature.}
    \label{fig:eff_vs_T}
\end{figure}

The single stage extraction efficiency is also studied at different temperatures. The Erlenmeyer flask is heated to the set temperature by water bath heating, then it is taken out and placed on a magnetic stirrer with heating function. The stirring bottle is equipped with thermal insulation cotton to keep it warm, and the temperature is monitored with an infrared thermometer. The stirring speed is 1200 rpm, and the time is 10 minutes. Fig.~\ref{fig:eff_vs_T} shows the relation between extraction efficiency and temperature. The extraction efficiency increases with temperature steadily. At 80 $^\circ$C, the efficiency reaches the level where the multi-stage extraction at room temperature ended up with. If the form of $^{212}$Pb in LS is not significantly different with increasing temperature, it indicates the single stage extraction is very efficiency in removing polar form of $^{212}$Pb at 80 $^\circ$C.

\subsection{Rotary disc contactor prototype}

\begin{figure}
    \centering
    \includegraphics[width=0.8\linewidth]{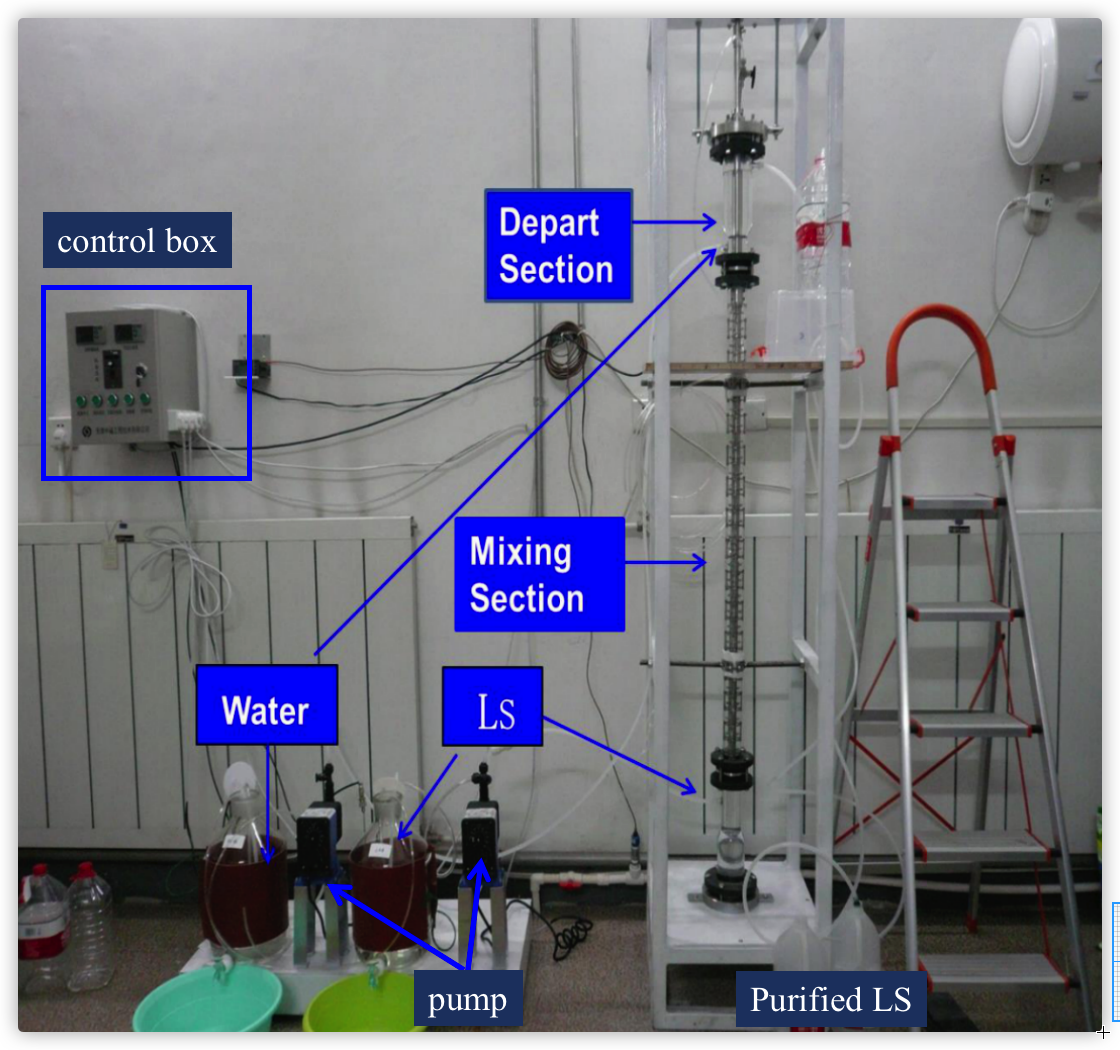}
    \caption{Picture of the RDC.}
    \label{fig:rdc_setup}
\end{figure}

A small prototype water extraction plant is constructed in the lab to understand larger system parameters. It is a rotating disk column (RDC), designed and manufactured by Institute of Process Engineering, CAS (Fig.~\ref{fig:rdc_setup}). It is designed to be operated in counter-current mode. The column is 3 meters high and 40 mm in diameter. It consists of 25 rotor stages with rotation speed up to 700 rpm. The working temperature can reach 80 $^{\circ}$C. The designed LS flow is 1.5 L/h, flow ratio between LS and water is 1:1 to 1:6 and this prototype can reach 3 theoretical stages in room temperature, 1 atm. The shaft of the prototype is made of polished 316L stainless steel, and the tower shell is made of acrylic. 

During operation, the phase fed from the bottom flows upwards is referred to as dispersed phase, while the one comes in from top of the tower and flows downwards is referred to as continuous phase. The stirring paddle disperses the dispersed phase into small droplets, and the small droplets travels through the continuous phase, when the mass transfer process occurs between two phases.

\subsubsection{Performance}
\label{sub:perfor}

\textbf{Rotation speed} An optimal rotation speed to form proper droplet size is crucial for good mass transfer. When the stirring speed is very low, the liquid scintillator and water are not sufficiently mixed. On the other hand, if the rotation speed is too high, the droplets will become smaller and fail to move up, which leads to flooding. Flooding is the phenomenon that the dispersed phase forms a continuous liquid column to prevent the extraction process. The extraction efficiency could maintain 80\% above while the rotation speed is over 500 rpm. At this time, the LS droplet diameter is from 1 to 2 mm.

\begin{figure}
    \centering
    \includegraphics[width=0.8\linewidth]{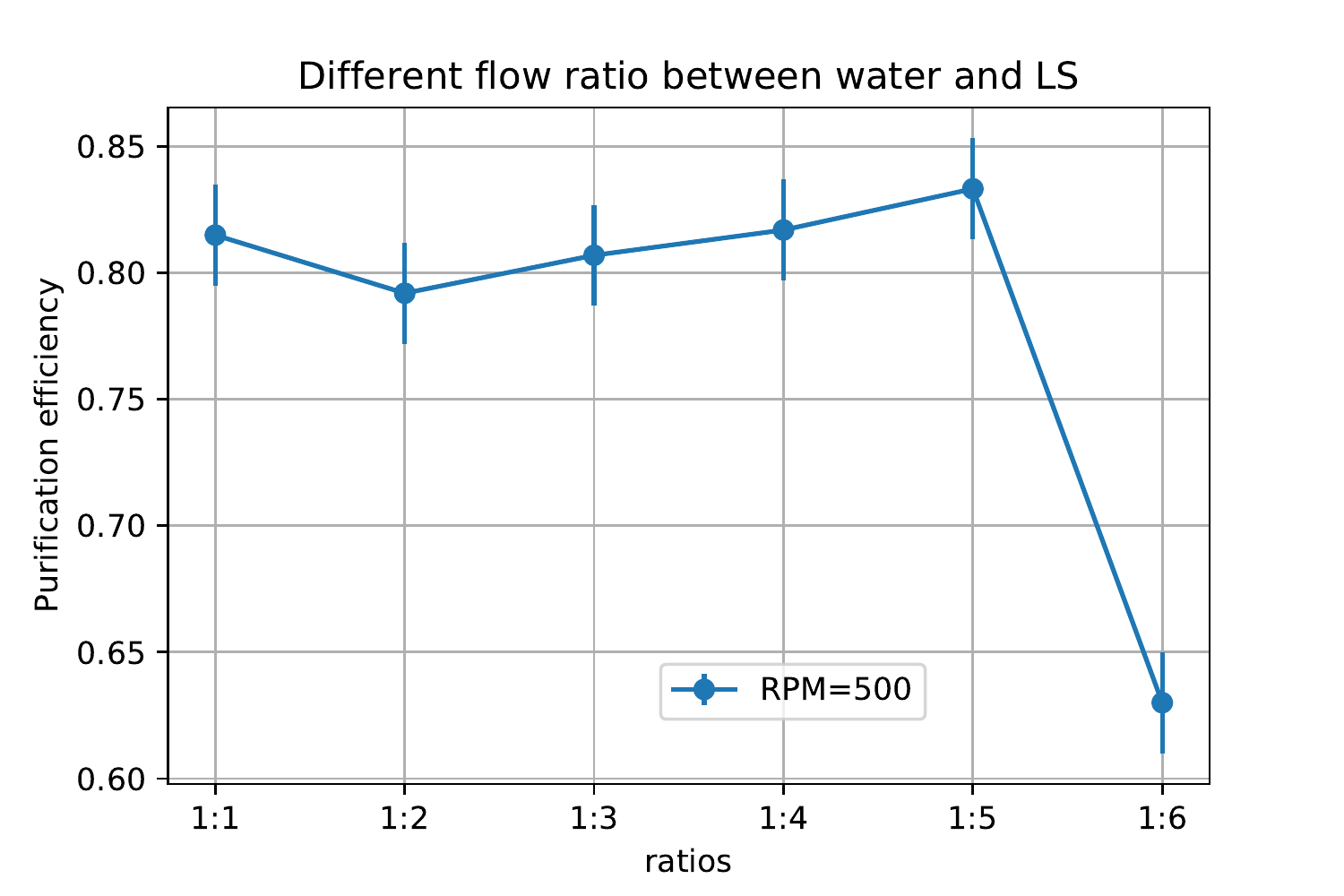}
    \caption{Purification efficiency of different flow ratios between water and LS}
    \label{fig:eff_vs_flow_ratios}
\end{figure}

\textbf{flow ratio} Based on Eq.~\eqref{eq:finaleff}, the flow rate ratio will affect the final extraction efficiency. Fig.~\ref{fig:eff_vs_flow_ratios} is the purification efficiency of different flow ratios between water and LS. The purification efficiency is studied at different flow ratio at the same rotation speed (500 RPM). We found the efficiency could maintain above 80\% for different flow ratio. However when flow ratio between water and LS is too small (i.e. smaller than 1:6), the efficiency drops dramatically. This indicates at 1:5, the equilibrium is already established, further increasing water does not improve the mass transfer.

\begin{figure}
    \centering
    \includegraphics[width=0.8\linewidth]{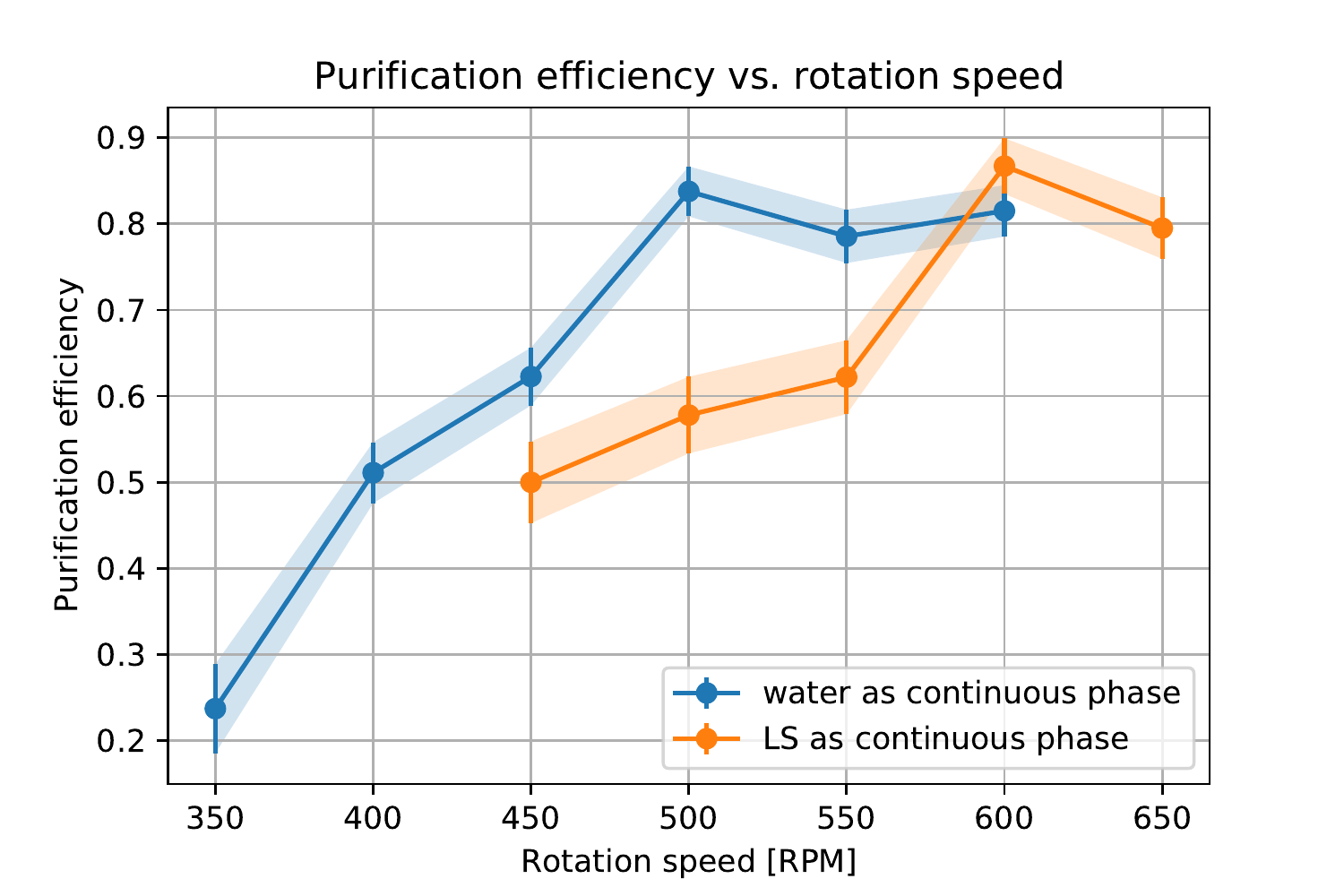}
    \caption{Purification efficiency of different continious phase}
    \label{fig:PDC}
\end{figure}

\textbf{Different continuous phase} The choice of whether LS or water should be used as dispersed phase is also studied. Fig.~\ref{fig:PDC} shows the purification efficiency curve with different rotation speed. When liquid scintillator is used as continuous phase, a relatively higher rotation speed is required to achieve an extraction efficiency of more than 80\%, and it is easier to flood with very small droplets, so that the entire extraction process is stopped. Besides, the entire extraction system is very unstable and flooding is easy to occur in LS as continuous phase.

\section{Daya Bay water extraction pilot plant}
\label{sec:dyb_plant}

We built a larger scale water extraction pilot plant at the Daya Bay experiment. The purpose of this pilot plant is that we use the 20 tons LS detector of Daya Bay experiment to check the LS quality produced by designed purify facilities, study the process parameters and gain engineering experience. Daya Bay experiment detectors distributed in 3 experiment halls (EH1, EH2 and EH3), we carried out pilot experiment using one of 20 tons LS detectors (AD1) in EH1 \cite{An:2015qga}. The original Gd-doped LS in the AD1 is completely pumped out and replaced with undoped LS coming out from the purification facilities \cite{Yan:2020bmc}. Daya Bay detector AD1 can detect \isou at 1$\times{10^{-15}}$g/g level. \isou content in the LS of AD1 is detected by the \isobi and \isopo cascade decay in the \isou chain. This detector can measure changes in \isou before and after purifying in different systems. Water extraction system is among one of them. The LS was self-circulated through purify facilities by turn to test the purification effects of each system.

\subsection{System description}
\begin{figure}
    \centering
    \includegraphics[width=0.8\linewidth]{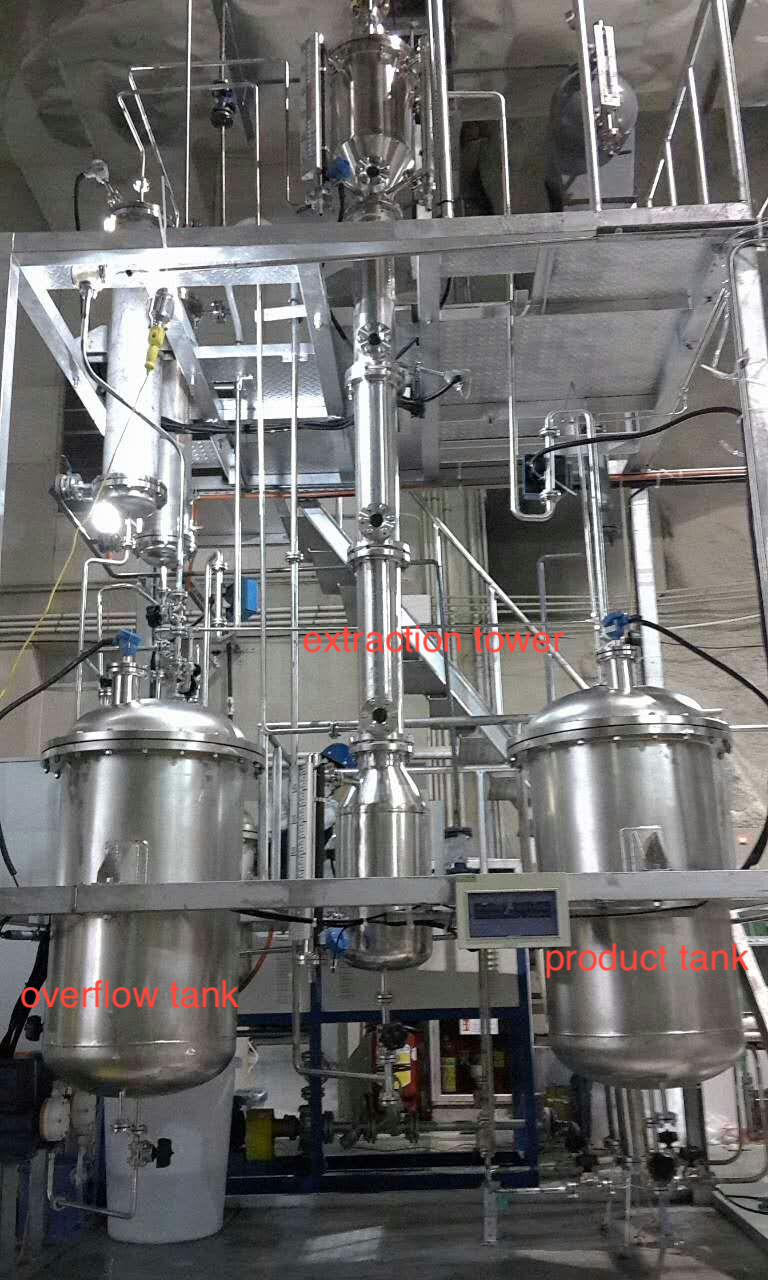}
    \caption{Picture of the Pilot Plant.}
    \label{fig:pilot plan}
\end{figure}

The role of water extraction as a part of the purification system is mainly to remove radioactive ions dissolved in liquid scintillator. The difference from the prototype is turbine extraction tower was adopted in water extraction pilot plant (Fig.~\ref{fig:pilot plan}). The extraction tower was designed and produced by Wuxi Hexiang Biochemistry Equipment Ltd. The extraction tower is 4.5m high and 0.15m in diameter. The flow rate of LS is 100 L/h and the flow rate of water is 30 L/h. Flow ratio of between LS and water is around 3:1. The working temperature is room temperature, working pressure is micro positive pressure. The paddle speed is 140 r/min. The droplet diameter is between 1 mm and 2 mm. The pilot plant extraction tower uses a specially designed turbine instead of the rotation disc of prototype to stir LS and water, which improves the stirring efficiency, reduces inter-stage flooding, and greatly improves the extraction efficiency.  The number of theoretical stages that the extraction tower can reach is greater than 5.  The system also includes 2 tanks, overflow tank and product tank. There are 100 nm PTFE filters before the LS entering the two tanks. 

In order to avoid that extraction plant polluting LS, tower, tanks and pipeline is made of 316L stainless steel. Inner surface is mechanical polished and electrochemical polished to roughness Ra$<$ 0.4 $\mu$ m. During assembling, tower components and the whole system were cleaned by the Alconox detergent to clean the grease and high pure water is used for flushing. Tower body and tank is chemical pickled and passivated to enhance its corrosion resistance. The same cleaning method and treatment method are also used for the pipeline. Orbital welding is adopted to avoid welding polluting LS.

To prevent Rn penetration, several methods were taken. The whole system was sealed and covered with high pure nitrogen. The dynamic shaft is sealed by magnetic force. Double O-rings flanges was used and inlet nitrogen between O-rings to get positive pressure. A liquid seal tank was set to prevent Rn permeation back from the vent.  To ensure the sealing, a leakage rate requirement was that partial flange connectors and valves less than 1$\times$10$^{-8}$ mbar$\cdot$ L/s, the whole system is less than 1$\times$10$^{-6}$ mbar$\cdot$ L/s. Airtight boxes were used to protect a part of valves and flow meters that could not meet our leakage rate requirement. During experiments, sealing of the water extraction system had met the design demands for Rn prevention.

\begin{figure}
    \centering
    \includegraphics[width=1\linewidth]{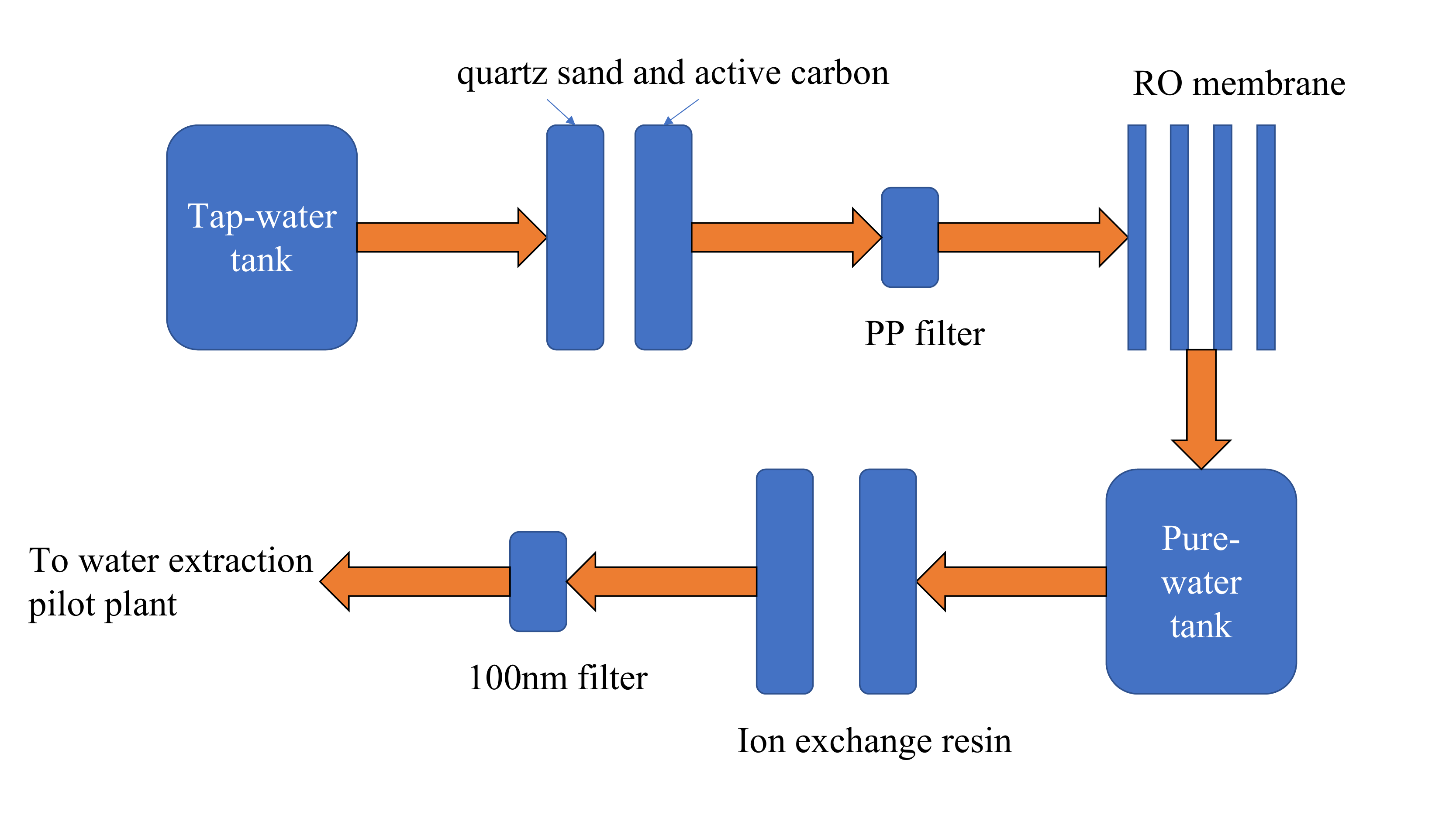}
    \caption{Pure water system of pilot plant}
    \label{fig:pws}
\end{figure}

\begin{figure}
    \centering
    \includegraphics{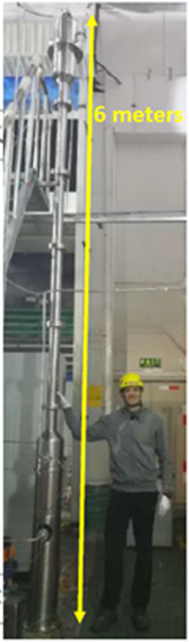}
    \includegraphics{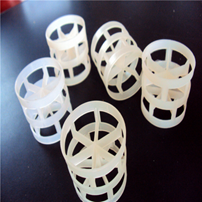}
    \caption{Nitrogen stripping column and Pall ring}
    \label{fig:Nitrogen stripping column}
\end{figure}

Since the water radioactivity level determines the radioactivity level of the final LS (eq.~\eqref{eq:finaleff}), the water used in extraction is supposed to be ultra-pure and have inherent radioactivity as less as possible. Fig.~\ref{fig:pws} is the pure water system for pilot plant. We purified tap water by the quartz sand filter, activated carbon, PP filter, two-stage RO membranes, ion exchange resin and 100nm filter in turn. Before the pure water enters the pilot plant, it will be stored in a storage tank, and the UV lamp will be irradiated for sterilization. The resistivity of final outlet water reached 18.2 $M\omega$. To remove Rn in water, we set up a nitrogen stripping column (Fig.~\ref{fig:Nitrogen stripping column}). Nitrogen stripping column is packed column, 6 meters total height, 80mm packing section diameter, the height of packed section was 3.5 m divided into three parts and each part was filled with 1 m high bulk packing and separated by Pall ring to avoid fluid short circuit. Packing is polypropylene bulk Pall ring with specific surface area 150 m$^2$/m$^3$. Water flow rate of nitrogen stripping system is 30 L/h and nitrogen flow rate is 2 m$^3$/h. 

The radon contents of water before nitrogen stripping was 25 Bq/m$^3$, measured by RAD-7. Radon contents of that after nitrogen stripping was measured indirectly by AD1. LS mixed with 30 L water per hour and we assumed that all the Rn in water would dissolve into LS, on the factual basis that the solubility of radon in LS is more than thirty times higher than that in water. Then the Rn concentration increase in AD1 was taken by water. According to the AD1 data, in the 5 days period of water extraction, Rn counts reduced slower than that decayed naturally. Rn contents of water was 20.1 mBq/m3, and Rn concentration was reduced 1200 times than before nitrogen stripping.

\begin{figure}
    \centering
    \includegraphics[width=0.9\linewidth]{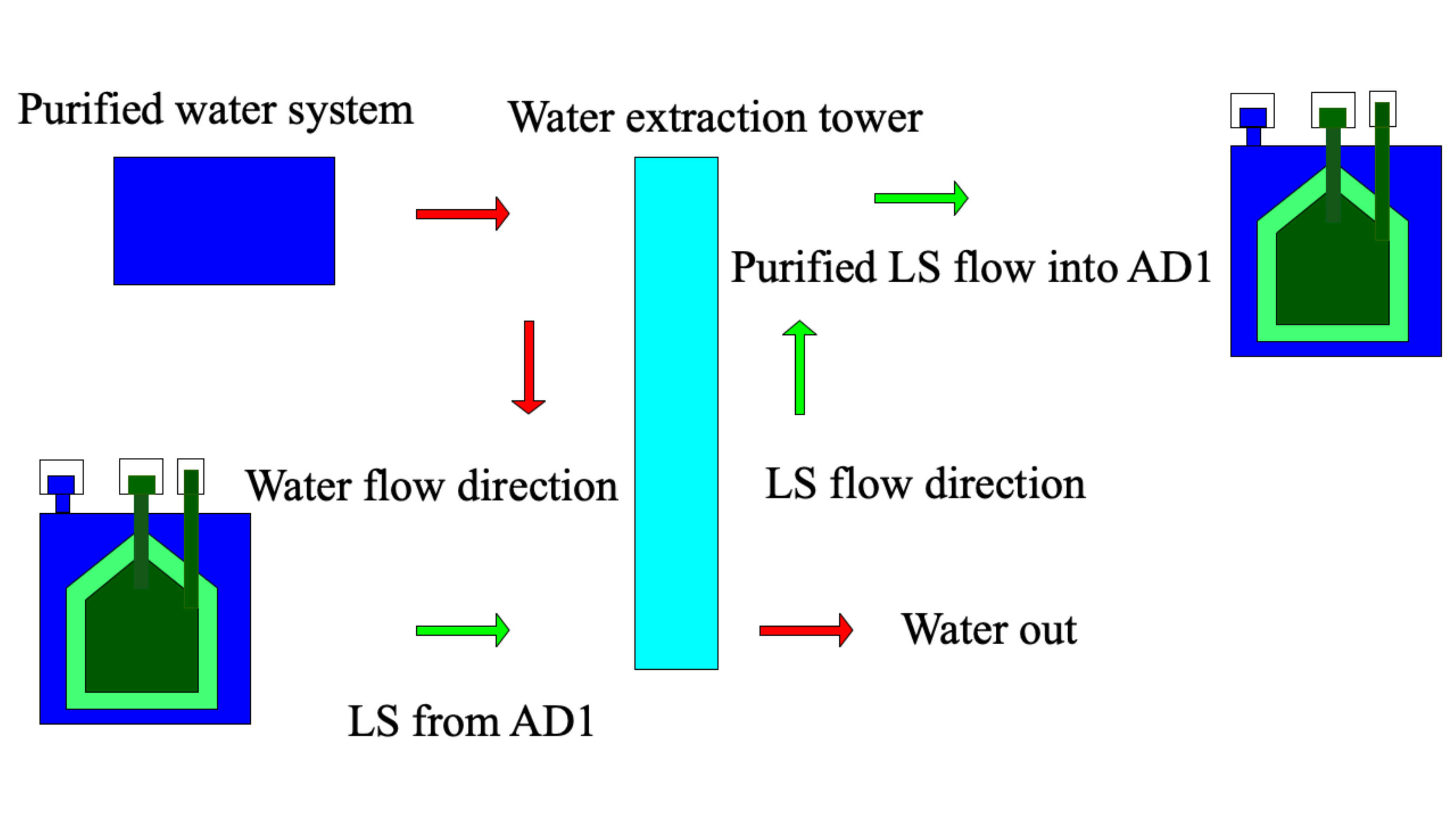}
    \caption{Water extraction of pilot plant flow-process diagram}
    \label{fig:pilot_plant_flow_process}
\end{figure}

Fig.~\ref{fig:pilot_plant_flow_process} is the water extraction pilot plant flow-process diagram. LS from AD1 fed from the bottom of extraction tower flows upwards, pure water flows downwards. The self-circulation will be carried out before the formal experiment. When the formal experiment is carried out, it will be as shown in the figure.
\subsection{Operation and Performance}

In pilot plant the water extraction system performed stable and some of the results had preliminary satisfied design demand: LS and water were agitated fully, droplet size 1$\sim$2 mm which was fit for mass transfer. In series running with other systems, $\isou$ content was purified to 2$\times10^{-15}$g$/$g. During the commissioning of the water extraction pilot plant, after the LS was purified and flow into product tank, the 100 nm filter was block. We had to change the filter a few times. We checked the replaced filter and found that there was a lot of water inside. We suspected that the water blocked the filter, making it unable to work normally under high LS flow. Therefore, in future full-scale water extraction plant, we decided to use nitrogen to bubbling in the LS to remove water. By contrasting the attenuation length(A.L.) of LS before and after the water extraction (Tab.~\ref{tbl1}), the A.L. did not obviously get worse, and it was considered that water extraction system did not pollute LS, and kept LS’s transparency. 

We want to know the individual purification efficiency of the water extraction system. When the water extraction system was operated alone, the uranium radioactivity level of the LS in AD1 did not drop further, nor did it rise significantly. The extraction efficiency did not meet expectations may be caused by a variety of reasons, like sealing problems, cleaning problems and high water radioactivity problems. If it is a sealing problem, it is easy to observe a significant increase in radioactivity in the AD1,  so this probability is low. Cleaning problems, such as tiny amount of dust entering the interior of the LS, will affect the radioactivity. Therefore, in the full-scale system, we will have more detailed cleaning steps and monitoring methods. Another likely problem is the radioactivity of the water. If the distribution coefficient of uranium is 5, the concentration of uranium in the water must be lower than 5$\times10^{-15}$g$/$g in order to make the concentration of uranium in the LS lower than 1$\times10^{-15}$g$/$g after extraction(eq.~\eqref{eq:finaleff}). Therefore, in future JUNO, the water will be purified separately to ensure that the water extraction can achieve the expected results. 

\begin{table}[width=.9\linewidth,cols=4,pos=h]
\caption{Attenuation length of different purification system. The error is 2.5 m}\label{tbl1}
\begin{tabular*}{\tblwidth}{@{} LL@{} }
\toprule
Sample name & A.L.(m)\\
\midrule
LAB after Al$_2$O$_3$ & 23.3 \\
LAB after distillation & 25.4 \\
LAB after water extraction & 24.6 \\
LAB after stream stipping & 24.4 \\
\bottomrule
\end{tabular*}
\end{table}

\section{Conclusion}
\label{sec:conc}
This article explores and measures a series of factors that may affect the final water extraction efficiency. The distribution coefficient of Pb with different temperature is measured. Multi-stage extraction efficiency is also measured in laboratory. Since the JUNO water extraction system will adopt counter-current extraction in the future, we built a prototype and a pilot plant to explore other factors. In the experiment of the prototype, we have given the change of the extraction efficiency with the rotation speed, and then determined the appropriate droplet diameter. The working mode of water as the continuous phase and liquid scintillation as the dispersed phase is determined. An appropriate flow rate ratio range is also studied. For the pilot test part, we obtained and shared a lot of experience in sealing protection(double O-rings, airtight boxes), cleaning, and filtering. Nitrogen stipping of water were proved to be highly effective. All of these will be valuable experience and references in the perspective of achieving low-background and high extraction efficiency in the final JUNO water extraction system.

\bibliography{cas-refs}

\end{document}